\documentclass[twocolumn,aps,floats,prb,showpacs]{revtex4-1}
\usepackage[latin9]{inputenc}
\usepackage{xcolor}
\usepackage{amsmath}
\usepackage{amssymb}
\usepackage{graphicx}

\makeatletter

\usepackage{color}
\usepackage{upgreek}
\usepackage{setspace}
\usepackage{pstricks}
\usepackage{dsfont}
\usepackage{fancyhdr}
\usepackage{array}
\usepackage{multirow}
\usepackage{placeins}

\newcommand{\eq}[1]{Eq.~(\ref{#1})}

\makeatother

\begin{document}

\title{Role of three-particle vertex within dual fermion calculations}

\author{T. Ribic$^{1,*}$, P. Gunacker$^{1,*}$, S. Iskakov$^{3,*}$,
M. Wallerberger$^{3}$, G. Rohringer$^{2}$, A. N. Rubtsov$^{2}$,
E. Gull$^{3}$ and K. Held$^{1}$}

\affiliation{$^{1}$Institute of Solid State Physics, TU Wien, 1040 Vienna, Austria
~\\
 $^{2}$Department of Physics, University of Michigan, Ann Arbor,
Michigan 48109, USA ~\\
 $^{3}$Russian Quantum Center, Novaya street, 100, Skolkovo, Moscow
region 143025, Russia } 

\date{\today}

\thanks{These authors contributed equally to this work}
\begin{abstract}
We investigate the influence of self-energy diagrams beyond the two-particle
vertex level within dual fermion theory. Specifically, we calculate
the local three-particle vertex and construct from it selected dual
fermion self-energy corrections to dynamical mean field theory. For
the two-dimensional Hubbard model, the thus obtained self-energy corrections
are small in the parameter space where dual fermion corrections based on the 
two-particle vertex only are small. However, in other parts of the parameter 
space, they are of a similar magnitude and qualitatively different from 
standard dual fermion theory. The high-frequency behavior of the self-energy 
correction is -- surprisingly -- even dominated by corrections stemming from
the three-particle vertex. 
\end{abstract}

\pacs{71.27.+a, 71.10.Fd , 71.30.+h}

\maketitle
\let\n=\nu \let\o=\omega \let\s=\sigma \global\long\def\Gred{\mathcal{G}}
 \global\long\def\Chired{\overset{\sim}{\chi}}


\section{Introduction}

\label{Sec:Intro} Strongly correlated electron systems pose some
of the greatest challenges in modern solid-state theory. The interplay
between the interaction that is diagonal in real space and the kinetic
energy that is diagonal in momentum space causes some fascinating,
albeit hard to describe physical phenomena. Analytical solutions to
interacting lattice fermion systems are scarce and numerical treatments
have to face the exponential growth of the Hilbert space with the
number of lattice sites. Quantum Monte Carlo methods, for their part,
suffer from the fermionic sign problem. In this situation, dynamical
mean field theory (DMFT)\citep{Metzner89a,MuellerHartmann89,Georges92a}
has become a standard method for the treatment of correlation effects
in fermionic lattice systems. By considering local correlations only,
DMFT self-consistently maps the lattice problem onto a single-site
Anderson impurity model. This model can be solved reliably by a variety
of algorithms. Often continuous-time quantum Monte Carlo (CT-QMC)
simulations \cite{Rubtsov2005,Werner2006,Gull2008a,Gull2011a} are employed to
this end because of their robustness, versatility, and the ability
to treat continuous baths. 

Nevertheless, DMFT is limited to {\em local} correlation effects
by construction. Hence, more recently, diagrammatic extensions of
DMFT have been at the focus of intense research efforts. These methods
aim to utilize the well-established local quantities derived from
DMFT as a starting point but add, on top of these, non-local correlations
 by means of Feynman diagrams. Examples of such diagrammatic
extensions of DMFT are the dynamical vertex approximation (D$\Gamma$A)\citep{DGAintro,DGA},
the dual fermion (DF)\citep{DF} theory and the one-particle-irreducible
approach (1PI)\citep{1PI} to mention just some of them; for a review see Ref.~\onlinecite{RMPVertex}.
A common feature of all diagrammatic extensions is that they build
upon the local (two- and more-particle) vertex and use it to construct non-local correlations in one- and
two-particle quantities. These approaches allow for describing physical
phenomena beyond the realm of DMFT, such as formation of a pseudogap
\cite{Katanin2009,Rubtsov2009,Jung2010,Jung2011,Taranto2014,Schaefer2015-2}
and (quantum) critical exponents \cite{Rohringer2011,Antipov2014,Hirschmeier2015,Schaefer2016}.

The mentioned diagrammatic extensions (D$\Gamma$A, DF and 1PI) should
-- in principle -- include local vertex functions up to infinite order
in the particle number. However, hitherto the application of these
theories has been mostly restricted to the two-particle level. On
the one hand it was argued that most of the relevant physics such
as spin fluctuations should already be included in diagrams generated
from the two-particle vertex (indeed in weak coupling perturbation
theory this physics is generated from similar diagrams with the bare
two-particle interaction instead of the vertex). On the other hand,
a very practical reason for the truncation at the level of the two-particle
vertex exists: three-particle vertices are numerically very expensive
to calculate and only recently enhanced computer resources and improved algorithms made such calculations feasible.
Furthermore, three-particle diagrammatics is much more complicated
to treat (also combinatorically) than two-particle diagrammatics.

To the best of our knowledge, there are only two previous papers that
include higher-order vertices within the DF framework. Ref.~\onlinecite{Hafermann2009}
found only weak effects of selected low-order diagrams on the leading
eigenvalue of the Bethe-Salpeter equation in the dual $ph$-channel
for the Hubbard model. In contrast, Ref.~\onlinecite{Ribic2017}
identified strong self-energy corrections due to the three-particle
vertex in the Falicov-Kimball model. 

It is the aim of this paper to further elucidate and to estimate the
influence of higher order vertex correlations on the self-energy within
DF. To this end, we calculate local three-particle vertices using
CT-QMC. From these we evaluate a simple self-energy diagram and investigate
its contribution in comparison to DMFT, the dynamical cluster approximation
(DCA), standard DF, 1PI and D$\Gamma$A. 

The study is conducted for the Hubbard model
with nearest neighbor hopping $t$ and interaction $U$ on a square
lattice which is described by the Hamiltonian 
\begin{equation}
{\mathcal{H}}=-t\sum_{\langle ij\rangle,\sigma}{c}_{i\sigma}^{\dagger}{c}_{j\sigma}^{\phantom{\dagger}}+U\sum_{i}{c}_{i\uparrow}^{\dagger}{c}_{i\uparrow}^{\phantom{\dagger}}{c}_{i\downarrow}^{\dagger}{c}_{i\downarrow}^{\phantom{\dagger}}\label{defhub}
\end{equation}
Here, $\langle ij\rangle$ denotes the summation over pairs of nearest
neighbor sites $i$ and $j$; and ${c}_{i\sigma}^{({\dagger})}$ annihilates
(creates) an electron on site $i$ with spin $\sigma$. In the following,
the half-bandwidth
($4t$) is chosen as the unit of energy, i.e., $4t\equiv1$.

The outline of the paper is as follows:
Section \ref{Sec:loal3P} is devoted to the calculation
of the local three-particle vertex. In Section
\ref{SubSec:3PGFVertex}, the form of this three-particle
vertex and how to obtain it from the three-particle Green's function by subtracting disconnected contributions and amputating Green's functions is discussed. The CT-QMC calculation of the three-particle
 Green's functions is outlined in turn in Section
\ref{SubSec:CTQMC}, with additional information  in Appendix
\ref{Sec:CTQMCappendix}.
The Feynman diagrams that we  consider in DF with this three-particle
vertex as a starting point and the corresponding equations are given
in Section
\ref{Sec:3PDF}. This is supplemented  in Appendix 
\ref{Sec:SDeq} by a derivation
of a generalized Schwinger-Dyson equation.
Section 
\ref{Sec:ResultsSigma} presents the results obtained for the two-dimensional Hubbard model. Finally, Section 
\ref{Sec:conclusion} provides a summary and an outlook.

\section{Calculation of local three-particle quantities within DMFT}
\label{Sec:loal3P}
\subsection{Three-particle Green's function and vertex}
\label{SubSec:3PGFVertex}
\begin{figure}
\includegraphics[width=0.7\linewidth]{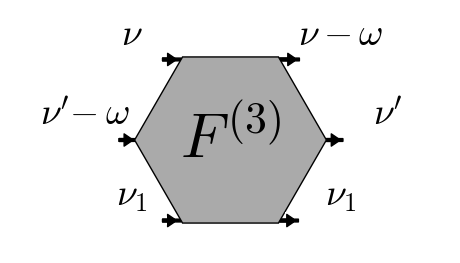} \caption{\label{3PVnotation} Frequency convention for the three-particle vertex
$F^{(3)}$. Note that we consider throughout the paper the situation
where no energy is transferred to frequency $\nu_{1}$ which reduces
the frequency dependence of $F^{(3)}$ to four frequencies. As for
spin degrees of freedom, $F^{\sigma_{1}\sigma_{2}\sigma_{3}}$ denotes
the vertex component where both lines associated with frequency $\nu_{1}$
carry the spin $\sigma_{1}$, the lines associated with $\nu$ carry
$\sigma_{2}$ and the lines associated with $\nu'$ carry $\sigma_{3}$. }
\end{figure}

Let us start by formally defining the local three-particle Green's function
\begin{multline}
G_{\nu_{1}\nu\nu'\omega}^{(3)\sigma_{1}\sigma_{2}\sigma_{3}}=\langle{c}_{\sigma_{1}}^{\dagger}(\nu_{1}){c}_{\sigma_{1}}(\nu_{1})\times \\
{c}_{\sigma_{2}}^{\dagger}(\nu-\omega){c}_{\sigma_{2}}(\nu){c}_{\sigma_{3}}^{\dagger}(\nu'){c}_{\sigma_{3}}(\nu'-\omega)\rangle,
\label{defg3}
\end{multline}
with three fermionic Matsubara frequencies $\nu_{1},\nu,\nu'$ and
one bosonic (transfer) frequency $\omega$, cf.~Appendix 
\ref{Sec:CTQMCappendix} for the Fourier-transformation from imaginary times. Fig. \ref{3PVnotation}
illustrates our frequency- and spin-convention for three-particle
quantities.

\begin{figure}
\includegraphics[width=1\linewidth]{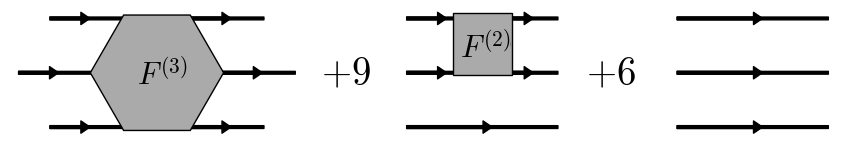} \caption{\label{3PVdecomp} The three-particle Green's function $G^{(3)}$
is decomposed into a three-particle connected contribution {[}$G_{C}^{(3)}=G^{(1)}G^{(1)}G^{(1)}F^{(3)}G^{(1)}G^{(1)}G^{(1)}${]},
nine terms consisting of a one-particle and connected two-particle
Green's functions {[}$G^{(1)}G_{C}^{(2)}${]}, and six disconnected
contributions {[}$G^{(1)}G^{(1)}G^{(1)}${]}. }
\end{figure}

To obtain the fully connected $n$-particle vertex functions
$F^{(n)}$ from $G^{(n)}$, first any disconnected contribution to
the propagators needs to be removed. Subsequently we need to amputate
the outer legs of the remaining, fully connected three-particle Green's
function $G_{C}^{(n)}$. On the two-particle level, there are only
two disconnected contributions to the Green's function $G^{(2)}$,
both consisting of a product of two one-particle Green's functions: $G^{(1)}G^{(1)}$.
On the three-particle level, there is much more variety among the
disconnected terms. A three-particle Green's function $G^{(3)}$ contains
terms disconnected into three one-particle propagators, $G^{(1)}G^{(1)}G^{(1)}$
(for example $\delta_{\omega,0}\ G_{\nu_{1}}^{(1)\sigma_{1}}G_{\nu}^{(1)\sigma_{2}}G_{\nu_{3}}^{(1)\sigma_{3}}$),
as well as other terms disconnected into a one-particle and a connected
two-particle Green's function, $G^{(1)}G_{C}^{(2)}$ (for example
$G_{\nu_{1}}^{(1)\sigma_{1}}G_{C\ \nu\nu'\omega}^{(2)\sigma_{2}\sigma_{3}}$),
as well as a fully connected term, see Fig. \ref{3PVdecomp} for an illustration.

\label{Sec:DMFT} 

\subsection{CT-QMC results for the local three-particle vertex functions}
\label{SubSec:CTQMC} 

\label{sec:NumResVert} Continuous-time quantum Monte Carlo (CT-QMC)
algorithms are based on a series expansion of the partition
function, and here employed for the Anderson impurity model. While
the specific Green's function measurement depends on the choice of
expansion, CT-QMC algorithms in general provide $n$-particle Green's
functions, consisting of fully connected as well as disconnected contributions.
Extracting irreducible vertex functions by subtracting disconnected
contributions and amputating outer legs, as discussed in the previous
Section, is a post-processing step to the simulation. CT-QMC algorithms
natively operate in the imaginary time domain. It is thus necessary
to define a suitable Fourier transform to recover the Matsubara frequency
representation of \eq{defg3}.

\begin{figure*}
\includegraphics[width=0.95\linewidth]{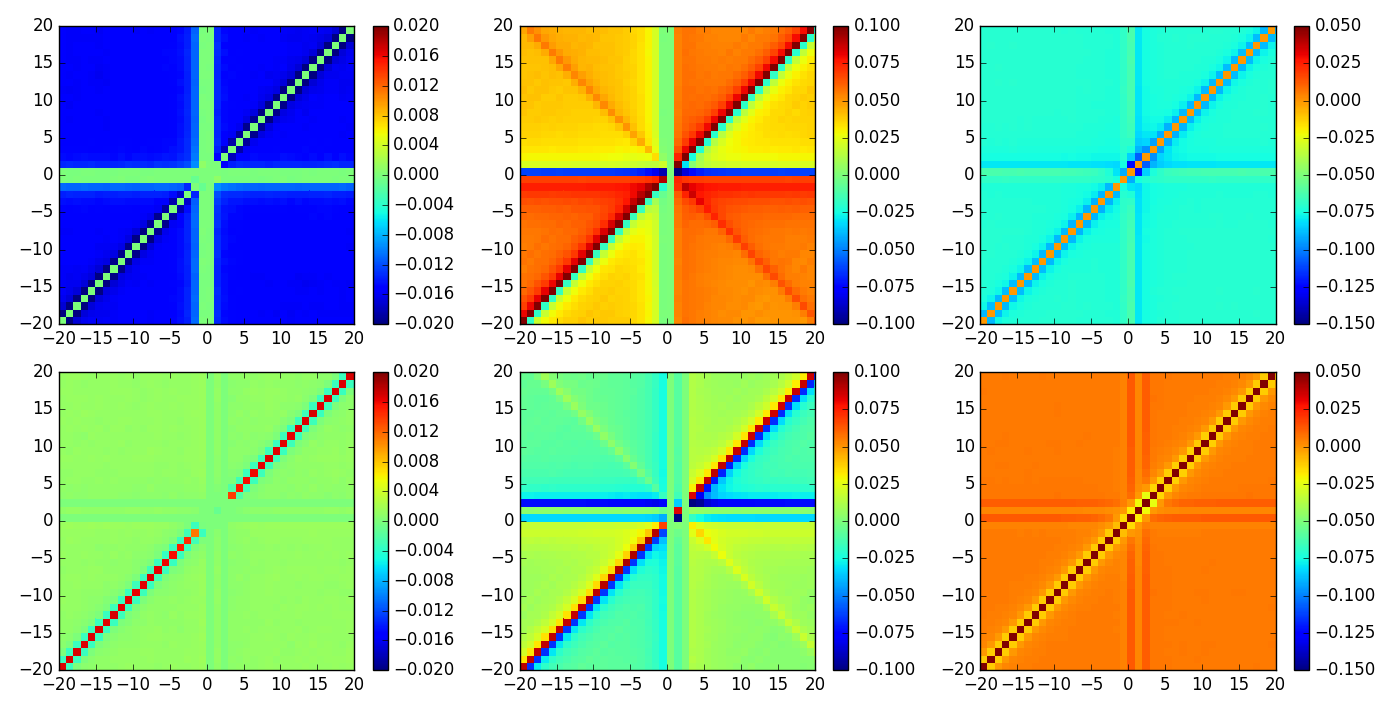} \caption{\label{AliceVertices} Full three-particle vertex $F^{\sigma_{1}\sigma\sigma'}$
(imaginary part) for the parameters $U=1,\beta=8$ and $n=1$. The frequency
$\nu_{1}$ was set to the first fermionic Matsubara frequency. The
heatplots are given as functions of Matsubara frequency $n_{\nu}$
and $n_{\nu}'$ {[}$\nu=(2n_{\nu}+1)\pi/\beta${]} for $\omega=0$
(upper row) and $\omega=4\pi/\beta$ (lower row) for the spin configurations
$\sigma_{1}\sigma\sigma'=\uparrow\uparrow\uparrow$ (left), $\sigma_{1}\sigma\sigma'=\uparrow\uparrow\downarrow$
(centre) and $\sigma_{1}\sigma\sigma'=\uparrow\downarrow\downarrow$
(right). }
\end{figure*}

Here, we calculate the three-particle Green's function for the auxiliary AIM associated to a DMFT solution at self-consistency, using both,
CT-QMC in the hybridization expansion (CT-HYB)\cite{Werner2006} and
in the auxiliary field expansion (CT-AUX).\cite{Gull2008} 
While in CT-AUX the single-particle
Green's function $G_{\nu}^{loc}$ is measured as a correction to the
non-interacting Green's function $G_{0}$, in CT-HYB the measurement
is achieved by cutting hybridization lines, not correcting any prior
Green's function object. In CT-AUX, corrections hence converge rapidly
in the high-frequency regions ($\sim1/\nu^{2}$), while the CT-HYB
result displays a constant error over the entire frequency range.
This becomes much more relevant for the vertex where we have,
as discussed, to amputate Green's function lines. This corresponds
to a division by a small number at large frequencies.  Hence the CT-HYB
three-particle vertex is noisy at larger frequencies, even more so
than the two-particle vertex. This makes weak-coupling CT-QMC
algorithms (e.g. CT-AUX)  more suitable for the calculation of the vertex
than strong-coupling algorithms (i.e. CT-HYB), at least when applied
to single-orbital models. However, we note that the high-frequency
behavior of CT-HYB algorithms is greatly alleviated by employing
improved estimators on the one-particle level \cite{Hafermann2012}
or vertex asymptotics on the two-particle level.\cite{Kaufmann2017}
Moreover, when eventually calculating the self-energy, the aforementioned small Green's functions are  multiplied again so that the noise at high-frequencies has 
a negligible effect for calculations based on the two-particle vertex.\cite{JPSJ-XXX} As we will see below this remains true for three-particle vertex corrections to the self-energy, but here only for the lowest few Matsubara frequencies.

of DMFT. 
Fig.~\ref{AliceVertices} shows the local three-particle
CT-AUX vertex calculated for the impurity problem of the DMFT solution
for the Hubbard-model at $U=1$, inverse temperature $\beta=8$,
and half-filling $n=1$. With the frequency $\nu_{1}$
fixed, the local three-particle vertex displays features very similar
to a two-particle vertex. A cross-like structure is visible along
the diagonal $\nu=\nu'$ and the secondary diagonal $\nu=-\nu'+\omega$.
A plus-like structure extends from $\nu=\pm \pi/\beta$ and $\nu'=\pm \pi/\beta$ as well as $\nu-\omega=\pm \pi/\beta$ and $\nu'-\omega=\pm \pi/\beta$. The features of the vertex are more pronounced for $\nu=+ \pi/\beta = \nu_1$,  $\nu'=+ \pi/\beta = \nu_1$ etc. than for $\nu=- \pi/\beta = -\nu_1$,  $\nu'=- \pi/\beta = -\nu_1$. Additionally, a constant background is present. The observed
structure is to be expected, as plotting the three-particle vertex
along the $\nu_{1}$ diagonal can be physically interpreted as the
scattering amplitude of a particle with energy $\nu_{1}$ with a particle
and a hole at energies $\nu,\nu'$ scattering with a transfer frequency
$\omega$.

\section{Dual fermion approach up to third order}
\label{Sec:3PDF}
\label{Sec:3PDF} 
The dual fermion approach allows for a systematic and -- in
principle -- exact decoupling of local and non-local degrees
of freedom for interacting lattice problems. This is achieved by a Hubbard-Stratonovich
transformation, which yields a so-called dual action of the form (see Ref.~\onlinecite{RMPVertex}):
\begin{multline}
\mathcal{S}_{dual}\left[\tilde{c}^{\dag};\tilde{c}^{\phantom{\dag}}\!\right]=\sum_{k_{1}}\dfrac{1}{G_{{\mathbf {k}}}-G_{\nu}^{loc}}\tilde{c}_{k_{1}}^{\dag}\tilde{c}_{k_{1}}^{\phantom{\dag}}+\\
\sum_{n=2}^{\infty}\sum_{{k_{1}},{k_{2}},{k_{3}},{k_{4}},...}\dfrac{1}{(n!)^{2}}F^{(n)}({k_{2}},{k_{1}},{k_{4}},{k_{3}},...)\tilde{c}_{k_{2}}^{\phantom{\dag}}\tilde{c}_{k_{1}}^{\dag}\tilde{c}_{k_{4}}^{\phantom{\dag}}\tilde{c}_{k_{3}}^{\dag}...\label{Eq:DFaction}
\end{multline}
Here, the Grassmann fields $\tilde{c}^{(\dag)}$ are associated with the
dual fermion degrees of freedom, and we use a four-vector plus spin
notation $k=({\mathbf{k}},\nu,\sigma)$. $G_{\nu}^{{\rm loc}}$ is
the local DMFT Green's function and $G_{k}$ the ${\mathbf {k}}$-dependent
DMFT Green's function for the Hubbard model that is obtained from
the Dyson equation and the local DMFT self-energy. The non-interacting
dual Green's function is given by $\widetilde{\mathcal{G}}_{0,k}=G_{{k}}-G_{\nu}^{{\rm loc}}$.
The full $n$-particle DMFT vertex functions $F^{(n)}$ are fully local and, hence, depend only
on the frequency- and spin-arguments and scatter
equally between all states obeying momentum conservation.

With the action \eq{Eq:DFaction} as a starting point we can calculate
via Feynman diagrammatic methods the interacting DF Green's
function $\widetilde{\mathcal{G}}_{k}$ and self-energy $\widetilde{\Sigma}_{k}$.
As we show in Appendix \ref{App:A}, the latter is connected to the
dual $n$-particle Green's function $\widetilde{\mathcal{G}}^{(n)}$
via a generalized Schwinger-Dyson equation (or Heisenberg equation
of motion) 
\begin{multline}
\widetilde{\Sigma}_{k}=-\sum_{n=2}^{\infty}\sum_{{k_{2}},{k_{3}},{k_{4}},...}\dfrac{(-1)^{n}}{n!(n-1)!}\\
F^{(n)}({k_{2}},k,{k_{4}},{k_{3}},...)\widetilde{\mathcal{G}}^{(n)}(k,{k_{2}},{k_{3}},{k_{4}},...)/\widetilde{\mathcal{G}}_{k}\label{SchwingerDysonMK2}
\end{multline}
Diagrammatically, the interpretation of the above equation is straightforward:
any dual self-energy diagram has to start with an interaction vertex.
Since there are infinitely many types of interaction vertices, an infinite sum of contributions to the self-energy exists.
Note that the dual Green's functions $\widetilde{\mathcal{G}}^{(n)}$
describe all possible diagrams which can be built from the original
local vertices $F^{(n)}$. The remaining external leg $\widetilde{\mathcal{G}}_{k}$
of the dual Green's function has to be amputated to generate a self-energy
diagram.

In \eq{SchwingerDysonMK2} full dual $n$-particle Green's functions appear. (not connected ones) However, any disconnected contribution to the Green's
function where a dual one-particle Green's function closes a loop
locally does not influence the dual self-energy if the one-particle
dual Green's functions are required to be completely non-local, i.e.\ $\sum_{\mathbf{k}}\widetilde{\mathcal{G}}_{\mathbf{k}\nu\sigma}=0$. For this reason, e.g.\ no Hartree or Fock term appears for the dual
fermions when truncating on the two-particle vertex level.

\begin{figure}
\begin{centering}
\includegraphics[width=0.75\linewidth]{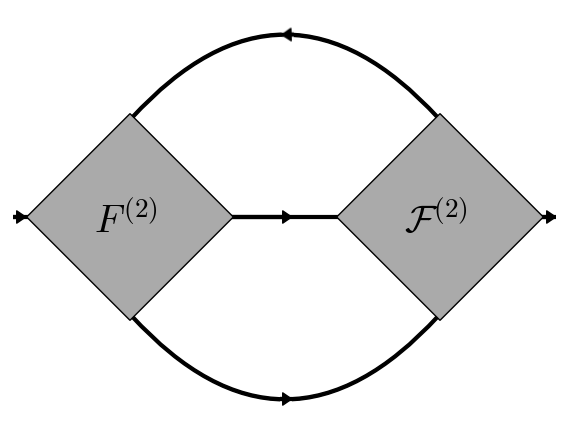} 
\par\end{centering}
\caption{\label{2ndCor} Feynman-diagrammatic representation of the dual self-energy
in terms of the local two-particle vertex $F^{(2)}$,
the dual propagator $\widetilde{\mathcal{G}}$ (line)
and the full DF vertex ${\cal F}^{(2)}$ (obtained, e.g., through
a ladder series). }
\end{figure}

In this paper, we consider local interaction terms up to the three-particle vertex in Eq.~(\ref{Eq:DFaction}). The actual choice of diagrams, which are constructed from these building blocks, is dictated by the physics of the system: In fact, for electrons on a bipartite lattice at (or close to) half-filling antiferromagnetic spin fluctuations are the predominant mechanism through which non-local correlations affect self-energies and spectral functions. Diagrammatically, such spin fluctuations are captured by ladder diagrams for $G^{(2)}$ (or equivalently $\widetilde{\mathcal{G}}^{(2)}$) in the $ph$ (and $\overline{ph}$) channel. Considering first Eq.~(\ref{SchwingerDysonMK2}) for $n=2$ we construct the diagram in Fig. \ref{2ndCor}. This ladder-based contribution to the dual self-energy corresponds to the standard choice for DF calculations in previous works.\citep{Brener2008,Hafermann2009}
The simplest contributions to $\widetilde{\mathcal{G}}^{(3)}$ in
Eq.~(\ref{SchwingerDysonMK2}) are the disconnected ones. For $\widetilde{\mathcal{G}}^{(3)}$,
an equivalent decomposition to the one in Fig. \ref{3PVdecomp} exists. In order to include antiferromagnetic spin fluctuations also for $n=3$ in Eq. \ref{SchwingerDysonMK2}, we consider the very same ladder diagrams for the disconnected contributions to $\widetilde{\mathcal{G}}^{(3)}$.
The terms of the form $\widetilde{\mathcal{G}}_{k}\widetilde{\mathcal{G}}_{k}\widetilde{\mathcal{G}}_{k}$ vanish for the same reason the Hartree-
and Fock terms vanish for the two-particle vertex: a closed Green's
function loop with $\sum_{\mathbf{k}}\widetilde{G}_{\mathbf{k}\nu\sigma}=0$.
The same holds for six out of the nine $\widetilde{\mathcal{G}}^{(2)}\widetilde{\mathcal{G}}_{k}$
terms contributing to $\widetilde{\mathcal{G}}^{(3)}$ analogously to the decomposition in Fig.~\ref{3PVdecomp}.
The remaining three possibilities contribute equally. Thus, within
our approximation, and taking into account all combinatorical prefactors our dual self-energy from the two- and three-particle vertex reads 
\begin{widetext}
\begin{multline}
\widetilde{\Sigma}_{k}\approx-\sum_{k_{2},k_{3},k_{4}}\dfrac{1}{2}F^{(2)}(k_{2},k,k_{4},k_{3})\ \widetilde{\mathcal{G}}_{0,k_{2}}\ \widetilde{\mathcal{G}}_{0,k_{3}}\ \widetilde{\mathcal{G}}_{0,k_{4}}\ \mathcal{F}^{(2)}(k,k_{2},k_{3},k_{4})\,\\
+\sum_{k_{1},k_{2},k_{3},k_{4}}\dfrac{1}{4}F^{(3)}(k,k,k_{2},k_{1},k_{4},k_{3})\ \widetilde{\mathcal{G}}_{0,k_{1}}\ \widetilde{\mathcal{G}}_{0,k_{2}}\ \widetilde{\mathcal{G}}_{0,k_{3}}\ \widetilde{\mathcal{G}}_{0,k_{4}}\ \mathcal{F}^{(2)}(k_{1},k_{2},k_{3},k_{4}).\label{Eq:DF23}
\end{multline}
\end{widetext}

\begin{figure}
\begin{centering}
\includegraphics[width=0.75\linewidth]{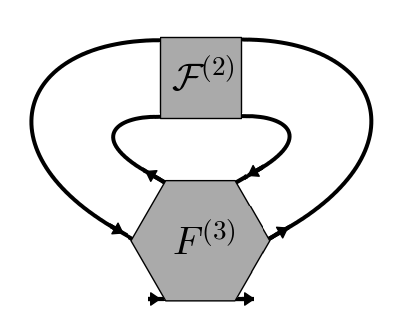} 
\par\end{centering}
\caption{\label{3rdCor} Feynman-diagrammatic representation of an additional
contribution to the dual self-energy that includes the local three-particle
vertex of the real fermions $F^{(3)}$. }
\end{figure}

The diagrammatic representation of the first line is given in Fig.
\ref{2ndCor}; it corresponds to standard DF and $n=2$ in \eq{SchwingerDysonMK2}.
The new contribution in the second line stems from $n=3$ and is illustrated
in Fig. \ref{3rdCor}. The vertex $\mathcal{F}^{(2)}$ in \eq{Eq:DF23}
denotes the full vertex of the dual fermions. In principle it can
be obtained from the action \eq{Eq:DFaction} or all Feynman diagrams
with ${F}^{(n)}$ and $\widetilde{\mathcal{G}}$ as building blocks.
Since an exact calculation of this quantity proves elusive, further
approximations are needed on its part. We employ the standard approximation
to this end, the $ph$ ladder approximation for $\mathcal{F}^{(2)}$:
\begin{equation}
\mathcal{F}_{kk'q, lad}^{(2)}=F_{kk'q}^{(2)}-\sum_{k_{1}}\mathcal{F}_{kk_{1}q, lad}^{(2)}\widetilde{\mathcal{G}}_{k_{1}}\widetilde{\mathcal{G}}_{k_{1}-q}F_{k_{1}k'q}^{(2)}.
\end{equation}
Where a three-variable notation 
\begin{equation}
\mathcal{F}_{kk'q}^{(2)}=\mathcal{F}^{(2)}(k,k-q,k'-q,k')
\end{equation}
was adapted. 

The self-energy as obtained in Eq. (\ref{Eq:DF23}) is the one for the dual electrons, i.e., it corrects the dual non-interacting Green's function. In order to obtain from it nonlocal correlations for real electrons it has to be transformed to the space of the original particles. For this purpose, the formalism of the DF theory provides an exact relation \cite{DF} which reads
\begin{equation}
\Sigma_{k}=\Sigma_{\nu}^{loc}+\dfrac{\widetilde{\Sigma}_{k}}{1+G_{\nu}^{loc}\widetilde{\Sigma}_{k}}.\label{Dualmap}
\end{equation}
While this relation certainly holds for the exact $\tilde{\Sigma}$ (i.e., where all diagrams for vertices of all orders are taken into account) it has been argued on the basis of diagrammatic considerations at weak coupling\cite{sixpoint, 1PI, RMPVertex} that it should be modified if only certain subsets of diagrams are considered: 
\begin{equation}
\Sigma_{k}=\Sigma_{\nu}^{{\rm loc}}+\widetilde{\Sigma}_{k},
\label{fullSigma}
\end{equation}
The (weak-coupling) arguments in favor of Eq. (9) given in the aforementioned Refs.~\onlinecite{1PI, RMPVertex} are also valid for the choice of diagrams for $\tilde{\Sigma}$ of the present paper. However, as there is no conclusive understanding regarding the choice of Eq. (8) or Eq. (9) for all coupling regimes, and an analysis of the difference between them are outside of the scope of this paper, we will consider both for the presentation of our numerical results in the next section.

\FloatBarrier

\vskip 5mm 

\section{Results: self-energy corrections}
\label{Sec:ResultsSigma}

\label{Sec:Sigma} 
\begin{figure*}
\includegraphics[width=0.95\linewidth]{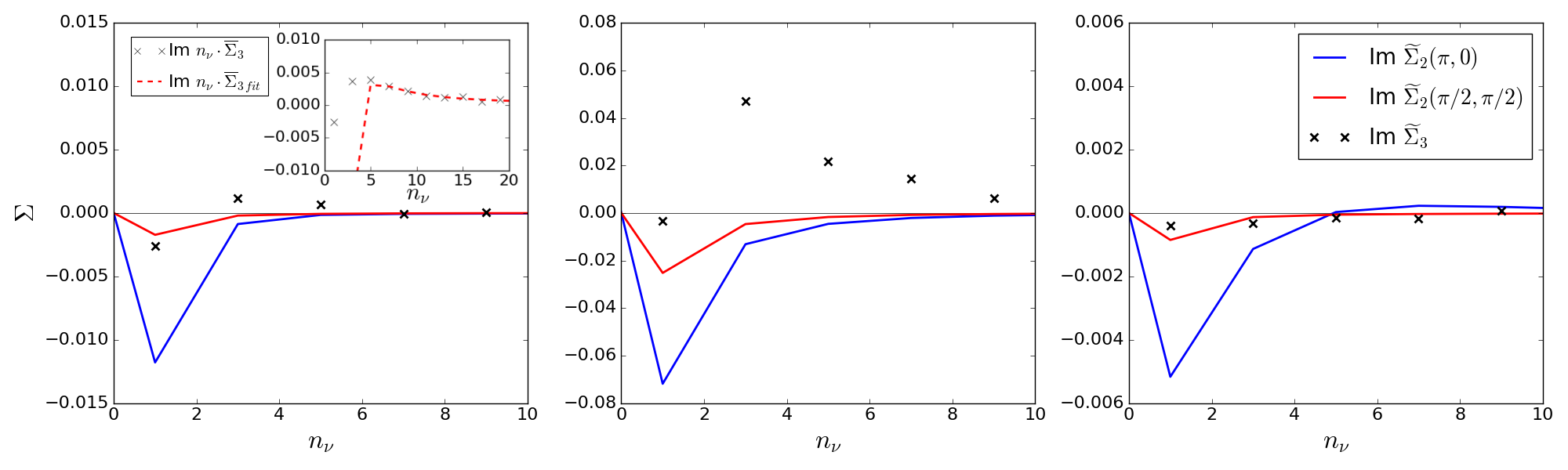}
\caption{\label{AliceSig2freq} Imaginary part of the dual self-energy 
correction of the standard DF theory $\widetilde{\Sigma}_{{\mathbf{k}}\nu}$
for two ${\mathbf k}$-points and the correction $\widetilde{\Sigma}_{3}$ based on the three particle
vertex and diagram Fig.~\ref{3rdCor}. From left to right, we present
data for $U=1$, $\beta=8$, $n=1$; $U=2$, $\beta=8$,
$n=1$ and $U=1$, $\beta=15$,
$n=0.8$, cf.~phase diagram Fig.~\ref{PD}. Inset in the first figure shows fitting function used to estimate high-frequency behavior.}
\end{figure*}

Let us now present the numerical results for one-shot
DF calculations based on converged DMFT baths for the two-dimensional
Hubbard model. For every discussed point, Fig.~\ref{AliceSig2freq}
shows the (Matsubara) frequency dependence of the DF self-energy correction
\footnote{The presented DF results are without self-consistency. However, for the parameters considered, imposing an inner self-energy self-consistency (not shown) leads only to minor modifications for $U=1$ and reduces both two- and three-particle corrections to about half their values for $U=2$. A closer investigation should also include an outer self-consistency with an update of the vertex and local problem, but is beyond the scope of the present paper.}
.
This self-energy needs to be
added to the DMFT self-energy to obtain the physical self-energy of
the Hubbard model. We compare in Fig.~\ref{AliceSig2freq} the standard
DF self-energy $\widetilde{\Sigma}_{2{\mathbf{k}}\nu}$ {[}first line
of \eq{SchwingerDysonMK2}{]} at the nodal $(\pi/2,\pi/2)$ and antinodal
$(\pi,0)$ ${\mathbf{k}}$-point of the Fermi surface with the selected
additional contribution based on the three-particle vertex {[}second
line of \eq{SchwingerDysonMK2}{]}. This specific three-particle
correction couples the two-particle ladder diagrams with the three
particle vertex, see Fig.~\ref{3rdCor}, and is ${\mathbf {k}}$-independent.

\begin{figure*}
\includegraphics[width=0.95\linewidth]{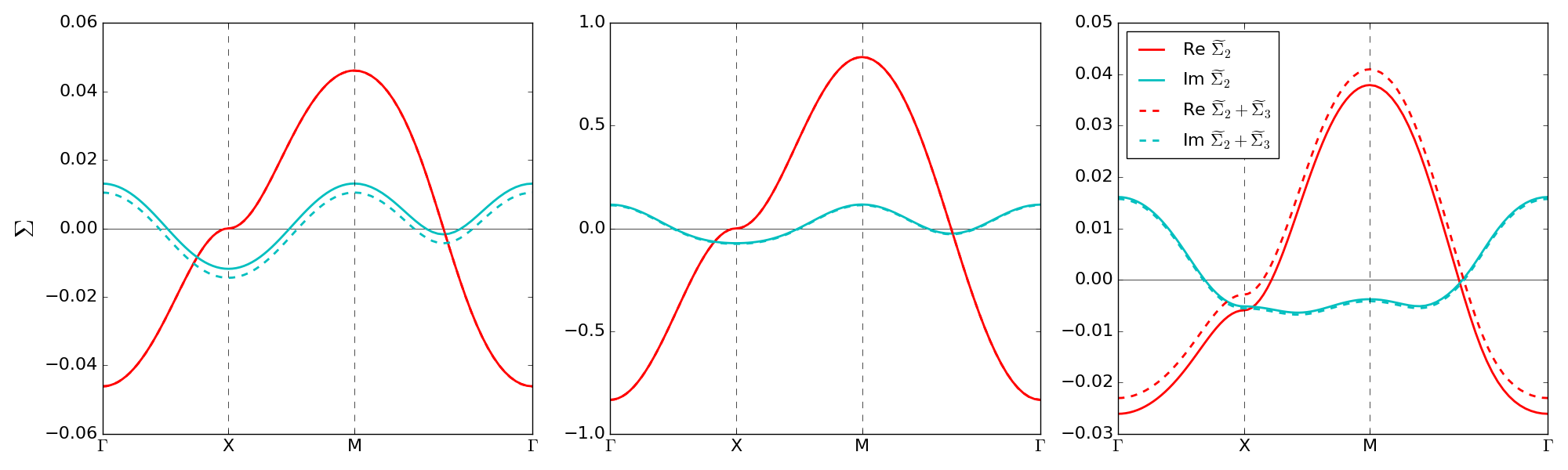}
\caption{\label{AliceSigBrillouin} Same as Fig.~\ref{AliceSig2freq} but
now presenting the ${\mathbf{k}}$-dependence of the DF self-energy
with ($\widetilde{\Sigma}_{2}+\widetilde{\Sigma}_{3}$) and without
($\widetilde{\Sigma}_{2}$) three-particle vertex corrections. The
figure shows the imaginary and real part of the DF self-energy at
the lowest Matsubara frequency along the path $\Gamma=(0,0)\rightarrow X=(\pi,0)\rightarrow M=(\pi,\pi)\rightarrow\Gamma$
through the Brillouin zone.}
\end{figure*}

Additionally, in Fig.~\ref{AliceSigBrillouin} the real and imaginary
part of the dual self-energy corrections is given along a path through
the Brillouin zone, including as well as excluding the three-particle
vertex correction. Because of its ${\mathbf {k}}$-independence, the
latter just gives a constant offset in these plots. Since the DF self-energy
is only a correction to the DMFT self-energy in \eq{fullSigma},
a positive imaginary part only means that the finite life time (damping)
effect of DMFT is reduced. For all investigated points the physical self-energy
remains negative.

Let us now discuss and interpret these results. At high temperatures
{[}($U=1$, $\beta=8$, $n=1$) and ($U=2$,
$\beta=8$, $n=1$){]} and for the doped system {[}($U=1$,
$\beta=15$, $n=0.8$){]}, the standard dual Fermion self-energy $\widetilde{\Sigma}_{2}$
is only a relatively small correction to the DMFT self-energy {[}${\rm Im}\Sigma_{n_{\nu}=1}^{{\rm loc}}$=
$-0.14$, $-0.96$ and $-0.075$, respectively{]}.
For $U=1$, the DF corrections based on the
three-particle vertex $\widetilde{\Sigma}_{3}$ are again considerably
smaller than $\widetilde{\Sigma}_{2}$. Note that this does not hold
for all ${\mathbf{k}}$-points. For example, the scattering rate due
to ${\rm Im}\widetilde{\Sigma}_{3}$ is larger than for ${\rm Im}\widetilde{\Sigma}_{2}$
for ${\mathbf{k}}=(\pi/2,\pi/2)$. But $\widetilde{\Sigma}_{2}$ is
much larger for ${\mathbf{k}}=(\pi,0)$, and also in general the variation
of $\widetilde{\Sigma}_{2}$ with ${\mathbf{k}}$ is much larger than
$\widetilde{\Sigma}_{3}$. While the three-particle vertex corrections appear
small in Fig.~\ref{AliceSigBrillouin}, Fig.~\ref{AliceSig2freq}
reveals that $\widetilde{\Sigma}_{3}$ is actually comparable in magnitude
to $\widetilde{\Sigma}_{2}$ when taking the second (not the first)
Matsubara frequency into account. This is particularly true for ($U=2$,
$\beta=8$, $n=1$) which happens to have a particularly small
$\widetilde{\Sigma}_{3}$ at the lowest Matsubara frequency.

\begin{figure}
\includegraphics[width=0.95\linewidth]{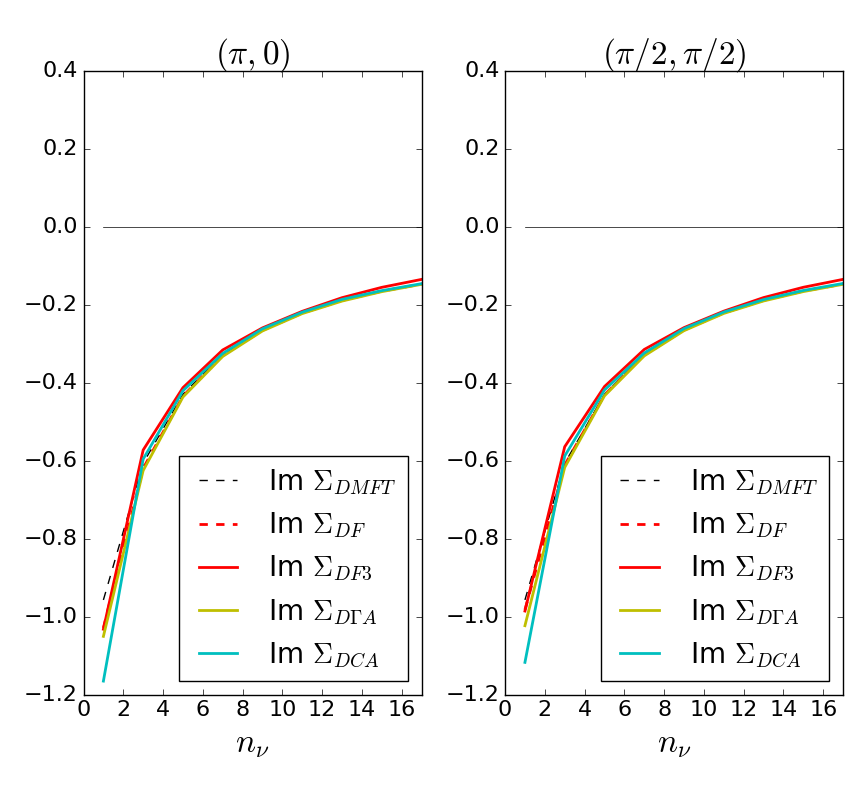} \caption{\label{MethodsComp} Comparison of the self-energy for the \textbf{{k}}-points
$(\pi,0)$ and $(\pi/2,\pi/2)$ calculated from DMFT, two-particle
dual fermion ($\Sigma_{DF}$), three-particle dual fermion ($\Sigma_{DF3}$),
D$\Gamma$A and DCA for $72$ lattice sites for the parameters $U=2$, $\beta=8$ and $n=1$. }
\end{figure}

We can trace these large DF contributions, both for $\widetilde{\Sigma}_{3}$
and $\widetilde{\Sigma}_{2}$, back to the strong enhancement of ${\cal F}^{2}$
in the ladder series for spin-$\uparrow\downarrow$ and ${\mathbf{q}}=(\pi,\pi)$.
Physically this corresponds to strong spin fluctuations in the two-dimensional
Hubbard model. For $\widetilde{\Sigma}_{2}$ these spin fluctuations
combine with one more interacting vertex $F^{(2)}$ in \eq{SchwingerDysonMK2}
to yield a strongly ${\mathbf{k}}$-dependent self-energy and pseudogap
physics. But the very same spin fluctuations also couple to the three-particle
vertex in \eq{SchwingerDysonMK2}, and yield a ${\mathbf{k}}$-independent
imaginary part of the self-energy of similar magnitude. We additionally compared the self-energies, as extracted from DMFT, dual fermion, based on two and three-particle vertices, D$\Gamma$A and DCA in Fig. \ref{MethodsComp}. The results were obtained $\beta = 8$.
The general trend, however, of the dominant fluctuations influencing the three-particle corrections in a sizable fashion is expected to persist within a stable, self-consistent approach.

An important remark is in order regarding the $1/i\nu$ asymptotic behavior of the self-energy $\Sigma_k$ for the real electrons: The correction $\tilde{\Sigma}^{(3)}$ [second line in Eq.(\ref{Eq:DF23})] gives rise to a $1/\nu$ contribution in $\tilde{\Sigma}$ (see inset in Fig. \ref{AliceSig2freq}, left panel). This modifies the -- already correct -- $1/i\nu$ asymptotics of the local DMFT self-energy and leads, hence, to a wrong $1/i\nu$ behavior of the total self-energy in Eqs. (\ref{Dualmap}) or (\ref{fullSigma}). Such a violation of the asymptotic behavior of the self-energy can be also observed in the D$\Gamma$A and the 1PI approaches \cite{Katanin2009,RohringerToschi2016, 1PI} and can be traced back \cite{RohringerToschi2016} to a violation of the Pauli principle at the two-particle level [i.e., more precisely to a violation of the sum rule $\frac{1}{\beta}\sum_{k} \chi_{\uparrow\uparrow}^{k}=\frac{n}{2}\left(1-\frac{n}{2}\right)$] in ladder based approaches. In the D$\Gamma$A and the 1PI approach this problem has been overcome \cite{Katanin2009, 1PI, RohringerToschi2016}) by renormalizing the corresponding spin- and/or charge-susceptibilities through a  Moriya $\lambda$ correction. Such a procedure could be also applied for the situation in this paper where the violation of the asymptotics of $\Sigma$ originates from the inclusion of the local three-particle vertex. An alternative route, which is more in the spirit of the DF method, would be to choose an appropriate (outer) self-consistency condition for the local reference system which removes the spurious asymptotic behavior. The question of which of the proposed methods is more suitable needs further discussions and goes beyond the scope of the present paper.

\begin{figure}
\includegraphics[width=0.95\linewidth]{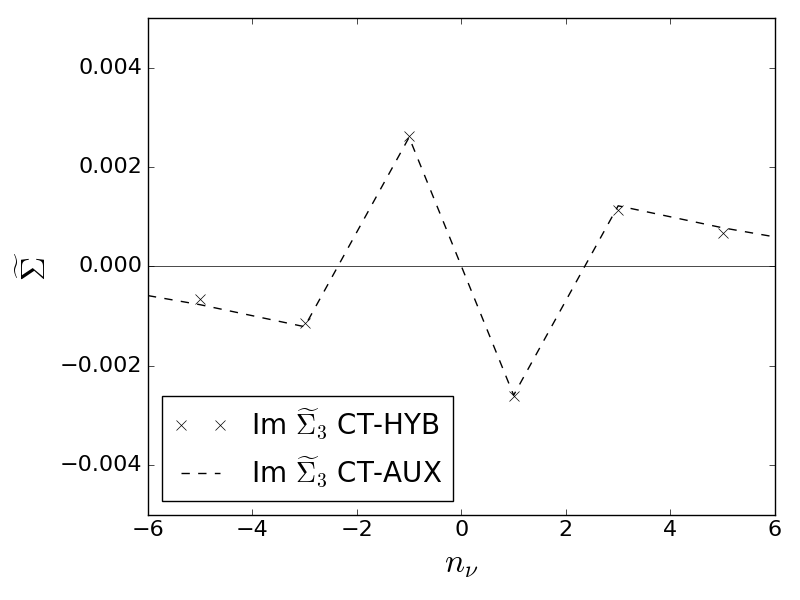} \caption{\label{HybIntCompare} Comparison of $\overline{\Sigma}_{3}$ calculated
from CT-AUX (dashed line) and CT-HYB (crosses) for the parameters $U=1$, $\beta=8$ and $n=1$.}
\end{figure}

Let us note that we find good agreement between DF calculations based
on vertices from CT-AUX and CT-HYB calculations, as exemplarily shown
in Fig. \ref{HybIntCompare}, though a separate investigation of the
vertices themselves showed that CT-AUX vertices display less noise,
especially at high frequencies. Let us note that for higher frequencies, outside the range of  Fig. \ref{HybIntCompare}, the  CT-HYB self-energy becomes more noisy.

\begin{figure}
\includegraphics[width=0.95\linewidth]{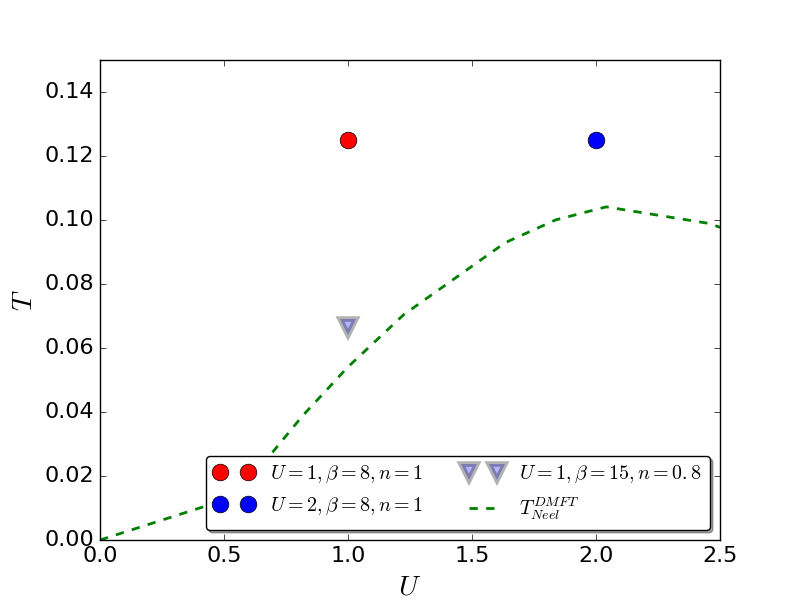} \caption{\label{PD} Positions of investigated points in parameter space relative to the N\'eel-temperature within DMFT for the half-filled system (which is an indication where spin fluctuations become more relevant). }
\end{figure}





\section{Conclusion}
\label{Sec:conclusion}

\label{Conclusio} We have calculated local three-particle Green's
functions and vertices employing CT-QMC algorithms in the hybridization
(CT-INT) and the auxiliary field expansion (CT-AUX). The structure
of the vertices for a fixed entering and leaving frequency $\nu_{1}$
is found to be similar to the two-particle case. High frequency features
persist in the Green's function, and by extension, the vertex functions.
Unavoidable noise in the high-frequency parts of the vertices has
only weak effects when calculating three-particle self-energy corrections
at small frequencies as the dual propagators within DF introduce enough
damping. For larger frequencies however, the high noise level of the
CT-INT vertex also reflects in a noisy self-energy, whereas the CT-AUX
vertex and constructed self-energy have a low statistical error.

For different points in the parameter space of the Hubbard model,
we find sizable corrections to the DF self-energy when including specific
three-particle diagrams. For high enough temperatures and for the
doped model, these three-particle vertex corrections are considerably
smaller than the standard DF self-energy. In particular they are smaller
than the two-particle DF corrections for the nodal point $(\pi/2,\pi/2)$.
In this parameter regime, our calculations indicate a proper convergence
of the DF theory when going to higher orders in the expansions (from
the $n=2$- to the $n=3$-vertex).

For higher interaction values, this picture changes. Spin fluctuations are the dominant driving force influencing the self-energy on the two-particle level. The same kind
of strong two-particle ladder contributions (the same kind of spin
fluctuations) couple additionally via the three-particle vertex to
an additional self-energy correction. This correction term yields
an additional $\mathbf{k}$-independent contribution to the imaginary
part of the self-energy, which can be interpreted as additional scattering
at spin fluctuations. The considered three-particle vertex correction
term also gives a $1/\nu$ asymptotic behavior which is absent in
standard DF and calls for a closer investigation.

\section*{Acknowledgments}

The authors
want to express special thanks to Thomas Sch\"afer, who made numerical
D$\Gamma$A results\citep{HubbDGA} available for a comparison. The
plots were made using the matplotlib \cite{Matplotlib} plotting library
for python. The local three-particle Green's functions were measured
in w2dynamics\citep{w2d1,w2d2} (CT-HYB) and ALPS\citep{ALPSnew} (CT-AUX); self-consistent DF calculations
were performed independently and benchmarked against OpenDF\citep{OpenDF}. Financial
support is acknowledged from the European Research Council under the
European Union's Seventh Framework Program (FP/2007-2013)/ERC through
grant agreement n.\ 306447 (TR, KH). PG has been supported by the
Vienna Scientific Cluster (VSC) Research Center funded by the Austrian
Federal Ministry of Science, Research and Economy (bmwfw). EG and SI are supported by the Simons collaboration on the many-electron problem. A contribution from G.R. and A.R. was funded by the Russian Science Foundation grant 16-42-01057. The computational
results presented have been achieved using the VSC and computational resources provided by XSEDE grant no. TG-DMR130036.
\FloatBarrier

\appendix

\section{CT-QMC measurement of the three-particle Green's function}
\label{Sec:CTQMCappendix}

For completeness, this Appendix attempts to briefly summarize the
construction of an estimator for the impurity three-particle Green's
function $G^{(3)}$ defined in Eq.~(\ref{defg3}). This is not a
comprehensive introduction, rather it can be read as an addendum to
Ref.~\onlinecite{Gull2011a}.

In the hybridization expansion (CT-HYB), we can define the interacting
Green's function by cutting hybridization lines from a given partition
function configuration, i.e.:
\begin{align}
G_{ij}(\tau,\tau^{\prime}) & = \frac{1}{Z} \frac{\delta Z}{\delta\Delta_{ji}(\tau^{\prime},\tau)}\nonumber \\
 & =\Big\langle\sum_{\alpha\beta}M_{\beta\alpha}\delta(\tau-\tau_{\alpha})\delta(\tau^{\prime}-\tau_{\beta}^{\prime})\delta_{ii_{\alpha}}\delta_{jj_{\beta}}\Big\rangle\nonumber \\
 & =:\langle g_{ij}(\tau,\tau^{\prime})\rangle\label{eq:z}
\end{align}
where $\Delta(\tau,\tau^{\prime})$ is the hybridization function,
$M_{\alpha\beta}^{-1}=\Delta_{i_{\alpha}j_{\beta}}(\tau_{\alpha},\tau_{\beta}^{\prime})$
is the matrix of hybridization lines, $\alpha$ and $\beta$ are indices
that run over the local creation and annihilation operators, respectively,
$i$ and $j$ denote spin-orbitals, and $\langle\cdot\rangle$ denote
the Monte Carlo sum over the configurations of $Z$. We introduce
the shorthand $g(\tau,\tau^{\prime})$ for the sum of all contributions
to the Green's function for a single configuration. Note that while
the expectation value $G(\tau,\tau^{\prime})$ is time-translational
invariant, this is not the case for the individual configuration $g(\tau,\tau^{\prime})$,
since the inner time indices of the diagram have not yet been integrated
over.

Generalizing Eq.~(\ref{eq:z}) to the three-particle Green's function,
we find:
\begin{align}
G_{ijklmn}(\tau_{1},\ldots,\tau_{6}) = & \langle g_{ij}(\tau_{1,}\tau_{2})g_{kl}(\tau_{3,}\tau_{4})g_{mn}(\tau_{5},\tau_{6})\nonumber \\
 & -g_{il}(\tau_{1,}\tau_{4})g_{kj}(\tau_{3,}\tau_{2})g_{mn}(\tau_{5},\tau_{6})\nonumber \\
 & -g_{in}(\tau_{1,}\tau_{6})g_{kl}(\tau_{3,}\tau_{4})g_{mj}(\tau_{5},\tau_{2})\nonumber \\
 & +g_{il}(\tau_{1,}\tau_{4})g_{kn}(\tau_{3,}\tau_{6})g_{mj}(\tau_{5},\tau_{2})\nonumber \\
 & +g_{in}(\tau_{1,}\tau_{6})g_{kj}(\tau_{3,}\tau_{2})g_{ml}(\tau_{5},\tau_{4})\nonumber \\
 & -g_{ij}(\tau_{1,}\tau_{2})g_{kn}(\tau_{3,}\tau_{6})g_{ml}(\tau_{5},\tau_{4})\rangle \; .\label{eq:g6wick}
\end{align}
This is nothing but the antisymmetrized sum over all possible removals
of three hybridization lines, which reflects the fact that Wick's
theorem is valid for the (non-interacting) bath propagator. The frequency
convention chosen in Eq.~(\ref{defg3}) translates to the following
definition of the Fourier transform:
\begin{align}
 & G_{\nu_{1}\nu\nu^{\prime}\omega}^{(3)\sigma_{1}\sigma_{2}\sigma_{3}}=\int_{0}^{\beta}d^{6}\tau\:G_{\sigma_{1}\sigma_{1}\sigma_{2}\sigma_{2}\sigma_{3}\sigma_{3}}(\tau_{1},\ldots,\tau_{6})\nonumber \\
 & \qquad\qquad\times\mathrm{e}^{\mathrm{i}(\nu_{1}\tau_{1}-\nu_{1}\tau_{2}+\nu\tau_{3}-(\nu-\omega)\tau_{4}+(\nu^{\prime}-\omega)\tau_{5}-\nu^{\prime}\tau_{6})}\label{eq:ft}
\end{align}
A naive implementation of Eq.~(\ref{eq:ft}) scales as $O(k^{6}N_{\omega}^{4})$,
where $k$ is the current expansion order and $N_{\omega}$ is the
number of frequencies, and is thus prohibitively expensive for even
moderate $k$. A binned measurement in imaginary time, while having
superior scaling $O(k^{6})$, is problematic, because $G_{ijklmn}(\tau_{1},\ldots,\tau_{6})$
is discontinuous on a set of hyperplanes $\tau_{i}=\tau_{j}$ and
their intersections, which in turn translate to large binning artefacts
in the Fourier transform. 

It is thus advantageous split the estimator into two parts: first,
we perform a Fourier transform of the single particle quantity from
Eq.~(\ref{eq:z})
\begin{align}
g_{ij}(\nu,\nu^{\prime}) & =\int_{0}^{\beta}d\tau\,d\tau^{\prime}\,g_{ij}(\tau,\tau^{\prime})\nonumber \\
 & =\sum_{\alpha\beta}M_{\beta\alpha}\exp(i\nu\tau_{\alpha}-\nu^{\prime}\tau_{\beta}^{\prime})\delta_{ii_{\alpha}}\delta_{jj_{\beta}},\label{eq:gnu}
\end{align}
which we can speed up by using a non-equidistant fast Fourier transform.
Note again that we need to retain both frequencies, as the quantity
is not time-translational invariant. Finally we perform the assembly
in Eq.~(\ref{eq:g6wick}) directly in Fourier space:
\begin{align}
 & G_{\nu_{1}\nu\nu^{\prime}\omega}^{(3)\sigma_{1}\sigma_{2}\sigma_{3}}=\langle g_{\sigma_{1}}(\nu_{1},\nu_{1})g_{\sigma_{2}}(\nu,\nu-\omega)g_{\sigma_{3}}(\nu^{\prime}-\omega,\nu^{\prime})\nonumber \\
 & \qquad-g_{\sigma_{1}}(\nu_{1},\nu-\omega)g_{\sigma_{2}}(\nu,\nu_{1})g_{\sigma_{3}}(\nu^{\prime}-\omega,\nu^{\prime})\delta_{\sigma_{1}\sigma_{2}}\nonumber \\
 & \qquad-g_{\sigma_{1}}(\nu_{1},\nu^{\prime})g_{\sigma_{2}}(\nu,\nu-\omega)g_{\sigma_{3}}(\nu^{\prime}-\omega,\nu_{1})\delta_{\sigma_{1}\sigma_{3}}\nonumber \\
 & \qquad+g_{\sigma_{1}}(\nu_{1},\nu-\omega)g_{\sigma_{2}}(\nu,\nu^{\prime})g_{\sigma_{3}}(\nu^{\prime}-\omega,\nu_{1})\delta_{\sigma_{1}\sigma_{2}\sigma_{3}}\nonumber \\
 & \qquad+g_{\sigma_{1}}(\nu_{1},\nu^{\prime})g_{\sigma_{2}}(\nu,\nu_{1})g_{\sigma_{3}}(\nu^{\prime}-\omega,\nu-\omega)\delta_{\sigma_{1}\sigma_{2}\sigma_{3}}\nonumber \\
 & \qquad-g_{\sigma_{1}}(\nu_{1},\nu_{1})g_{\sigma_{2}}(\nu,\nu^{\prime})g_{\sigma_{3}}(\nu^{\prime}-\omega,\nu-\omega)\delta_{\sigma_{2}\sigma_{3}}\rangle\; . \label{eq:g6wickiw}
\end{align}
Putting it all together, this reduces the scaling to $O(k^{2}N_{\omega}\log N_{\omega})+O(N_{\omega}^{4})$,
which improves also on the scaling of the time binning and makes the
estimator computationally feasible.

It is worth pointing out that Eq.~(\ref{eq:g6wick}), and in general
any estimator for $n>1$ particles constructed in this fashion, is
not valid for systems with interactions beyond density-density type
and a hybridization function that is (block-)diagonal in $i$ and
$j$. In such case, one would have to resort to worm sampling, which
we however gauge as a formidable computational challenge in itself
due to the sheer size of the worm configuration space and the size
of the measured object itself. Fortunately, this is not an issue here,
as we are studying the single-orbital case.

In the auxiliary field expansion (CT-AUX), one follows the same procedure
of applying Eq.~(\ref{eq:g6wickiw}) to a Fourier transformed quantity.
However, since the CT-AUX estimator is formed by adding a pair of
local operators rather than cutting hybridization lines, the single
particle contribution is instead given by:
\begin{align}
g_{\sigma}(\nu,\nu^{\prime}) & =G_{0\sigma}(\nu,\nu^{\prime})+G_{0\sigma}(\nu,\nu^{\prime})m_{\sigma}(\nu,\nu^{\prime})G_{0\sigma}(\nu,\nu^{\prime})\label{eq:gaux}\\
m_{\sigma}(\nu,\nu^{\prime}) & =\sum_{\alpha\beta}M_{\alpha\beta}\exp(i\nu\tau_{\alpha}-\nu^{\prime}\tau_{\beta}^{\prime})\delta_{\sigma\sigma_{\alpha}\sigma_{\beta}},\label{eq:maux}
\end{align}
where $G_{0}$ is the non-interacting Green's functions, and $M_{\alpha\beta}$
is the matrix of auxiliary spin system. The scaling for the estimator
is the same as for the CT-HYB case; however, it is evident from Eq.~(\ref{eq:gaux})
that the CT-AUX estimator is more well-behaved at large frequencies,
since the Monte Carlo signal drops as $1/\nu^{2}$ .

\section{Derivation of generalized Schwinger-Dyson equation}
\label{Sec:SDeq}
\label{App:A} Starting from the dual fermion action \eq{Eq:DFaction},
we can rewrite the functional-integral expression for the dual fermion
propagator $\widetilde{\mathcal{G}}$ as 
\begin{equation}
\widetilde{\mathcal{G}}_{k}=\dfrac{\int\mathcal{D}\left[\tilde{c}^{\dag}\tilde{c}^{\phantom{\dag}}\!\right]e^{\mathcal{S}_{dual}}\tilde{c}_{k}^{\dag}\tilde{c}_{k}^{\phantom{\dag}}}{\int\mathcal{D}\left[\tilde{c}^{\dag}\tilde{c}^{\phantom{\dag}}\!\right]e^{\mathcal{S}_{dual}}}.\label{FuncIntGdual}
\end{equation}
Let us now systematically decompose $\mathcal{S}_{dual}$ into two
parts $\mathcal{S}_{dual}^{k}$ and $\mathcal{S}_{dual}^{\neg k}$,
where $\mathcal{S}_{dual}^{k}$ consists of all summands containing
$\tilde{c}_{k}^{\dag}$ and $\mathcal{S}_{dual}^{\neg k}$ of all the remaining
ones (containing no $\tilde{c}_{k}^{\dag}$). Since all terms in the action
have an even number of Grassmann-fields, they commute and we can write
\begin{equation}
e^{\mathcal{S}_{dual}}=e^{\mathcal{S}_{dual}^{k}}e^{\mathcal{S}_{dual}^{\neg k}}.
\end{equation}
We also know that 
\begin{equation}
\left(\mathcal{S}_{dual}^{k}\right)^{2}=0,
\end{equation}
because all of its constituting terms contain $\tilde{c}_{k}^{\dag}$ (and
$(\tilde{c}_{k}^{\dag})^{2}=0$). Therefore, we also have 
\begin{equation}
e^{\mathcal{S}_{dual}^{k}}=\left(1+\mathcal{S}_{dual}^{k}\right)
\end{equation}
and 
\begin{equation}
\mathcal{S}_{dual}^{k}\cdot e^{\mathcal{S}_{dual}^{k}}=\mathcal{S}_{dual}^{k}.\label{GrassMann}
\end{equation}
We use the relations above to rewrite \eq{FuncIntGdual} as\footnote{In the denominator, the $1$ in the expansion $\exp^{\mathcal{S}_{dual}^{k}}=\left(1+\mathcal{S}_{dual}^{k}\right)$
vanishes when integrating. The remaining term $\mathcal{S}_{dual}^{k}$
can be rewritten according to \eq{GrassMann}. } 
\begin{equation}
\widetilde{\mathcal{G}}_{k}=\dfrac{\int\mathcal{D}\left[\tilde{c}^{\dag}\tilde{c}^{\phantom{\dag}}\!\right]e^{\mathcal{S}_{dual}}\tilde{c}_{k}^{\dag}\tilde{c}_{k}^{\phantom{\dag}}}{\int\mathcal{D}\left[\tilde{c}^{\dag}\tilde{c}^{\phantom{\dag}}\!\right]e^{\mathcal{S}_{dual}}\mathcal{S}_{dual}^{k}}.\label{FuncIntGdual2}
\end{equation}
The next steps in expressing the dual self-energy are a division of
both enumerator and denominator in \eq{FuncIntGdual2} by the dual
partition function $\int\mathcal{D}\left[\tilde{c}^{\dag}\tilde{c}^{\phantom{\dag}}\!\right]e^{\mathcal{S}_{dual}}$
and an explicit decomposition of $\mathcal{S}_{dual}^{k}$. 
\begin{multline}
\mathcal{S}_{dual}^{k}=\left(\widetilde{\mathcal{G}}_{0,k}\right)^{-1}\tilde{c}_{k}^{\dag}\tilde{c}_{k}^{\phantom{\dag}}+\\
\sum_{n=2}^{\infty}\sum_{k_{1},k_{2},k_{3},...}\dfrac{1}{n!(n-1)!}F^{(n)}(k_{1},k,k_{3},k_{2},...)\tilde{c}_{k_{1}}^{\phantom{\dag}}\tilde{c}_{k}^{\dag}\tilde{c}_{k_{3}}^{\phantom{\dag}}\tilde{c}_{k_{2}}^{\dag}...
\label{Eq:Sdual2}
\end{multline}
Here, the sum over all $k$ is gone as only the terms containing $\tilde{c}_{k}^{\dag}$
are included in $\mathcal{S}_{dual}^{k}$; multiple possibilities
for the summed-over indices to generate the index $k$ are taken care
of by replacing one of the factors $n!$ by $(n-1)!$. We now restore
normal ordering to the Grassmann-variables in $\mathcal{S}_{dual}^{k}$,
yielding another factor $(-1)^{n}$ for the term containing vertices. 
Inserting Eq.~(\ref{Eq:Sdual2}) into Eq.~(\ref{FuncIntGdual2}), we get
\begin{widetext}
\begin{equation}
\widetilde{\mathcal{G}}_{k}=\dfrac{\dfrac{\int\mathcal{D}\left[\tilde{c}^{\dag}\tilde{c}^{\phantom{\dag}}\!\right]e^{\mathcal{S}_{dual}}\tilde{c}_{k}^{\dag}\tilde{c}_{k}^{\phantom{\dag}}}{\int\mathcal{D}\left[\tilde{c}^{\dag}\tilde{c}^{\phantom{\dag}}\!\right]e^{\mathcal{S}_{dual}}}}{\dfrac{\int\mathcal{D}\left[\tilde{c}^{\dag}\tilde{c}^{\phantom{\dag}}\!\right]e^{\mathcal{S}_{dual}}(\widetilde{\mathcal{G}}_{0,k})^{-1}\tilde{c}_{k}^{\dag}\tilde{c}_{k}^{\phantom{\dag}}+\sum_{n=2}^{\infty}\sum_{k_{1},k_{2},k_{3},...}\dfrac{(-1)^{n}}{n!(n-1)!}F^{(n)}(k_{1},k,k_{3},k_{2},...)\tilde{c}_{k}^{\dag}\tilde{c}_{k_{1}}^{\phantom{\dag}}\tilde{c}_{k_{2}}^{\dag}\tilde{c}_{k_{3}}^{\phantom{\dag}}...}{\int\mathcal{D}\left[\tilde{c}^{\dag}\tilde{c}^{\phantom{\dag}}\!\right]e^{\mathcal{S}_{dual}}}}.\label{FuncIntGdual3}
\end{equation}
\end{widetext}

The enumerator by itself yields $\widetilde{\mathcal{G}}_{k}$ when performing
the Grassmann-integration, while in the denominator a sum of $n$-particle
dual Green's functions multiplied by $n$-particle DMFT vertex functions
and a term $\widetilde{\mathcal{G}}_{k}(\widetilde{\mathcal{G}}_{0,k})^{-1}$
appear. We divide both by the enumerator $\widetilde{\mathcal{G}}_{k}$, and end
up with 
\begin{multline}
\widetilde{\mathcal{G}}_{k}=\Bigg(\left(\widetilde{\mathcal{G}}_{0,k}\right)^{-1}+\sum_{n=2}^{\infty}\sum_{k_{1},k_{2},k_{3},...}\dfrac{(-1)^{n}}{n!(n-1)!}\\
F^{(n)}(k_{1},k,k_{3},k_{2},...)\widetilde{\mathcal{G}}^{(n)}(k,k_{1},k_{2},k_{3},...)/\widetilde{\mathcal{G}}_{k}\Bigg)^{-1},\label{Eq:Heisenberg}
\end{multline}
where $\widetilde{\mathcal{G}}^{(n)}$ denotes the dual $n$-particle
Green's function. Employing Dyson's equation, we recover an exact
expression for the self-energy of the dual fermions 
\begin{multline}
\widetilde{\Sigma}_{k}=-\sum_{n=2}^{\infty}\sum_{k_{1},k_{2},k_{3},...}\dfrac{(-1)^{n}}{n!(n-1)!}\\
F^{(n)}(k_{1},k,k_{3},k_{2},...)\widetilde{\mathcal{G}}^{(n)}(k,k_{1},k_{2},k_{3},...)/\widetilde{\mathcal{G}}_{k}\label{SchwingerDysonMK2_app}
\end{multline}
that is reminiscent of the Schwinger-Dyson equation.







\bibliographystyle{unsrt}
\bibliography{Refs} 

\begin{thebibliography}{10}

\bibitem{Metzner89a}
Walter Metzner and Dieter Vollhardt.
\newblock Correlated lattice fermions in $d=\infty$ dimensions.
\newblock {\em Phys. Rev. Lett.}, 62:324--327, Jan 1989.

\bibitem{MuellerHartmann89}
E.~M\"uller-Hartmann.
\newblock Correlated fermions on a lattice in high dimensions.
\newblock {\em Zeitschrift f\"ur Physik B Condensed Matter}, 74:507--512, 1988.

\bibitem{Georges92a}
Antoine Georges and Gabriel Kotliar.
\newblock Hubbard model in infinite dimensions.
\newblock {\em Phys. Rev. B}, 45:6479--6483, Mar 1992.

\bibitem{Rubtsov2005}
A.~N. Rubtsov, V.~V. Savkin, and A.~I. Lichtenstein.
\newblock Continuous-time quantum monte carlo method for fermions.
\newblock {\em Phys. Rev. B}, 72:035122, Jul 2005.

\bibitem{Werner2006}
Philipp Werner, Armin Comanac, Luca de' Medici, Matthias Troyer, and Andrew~J.
  Millis.
\newblock Continuous-time solver for quantum impurity models.
\newblock {\em Phys. Rev. Lett.}, 97:076405, Aug 2006.

\bibitem{Gull2008a}
E.~Gull, P.~Werner, O.~Parcollet, and M.~Troyer.
\newblock Continuous-time auxiliary-field monte carlo for quantum impurity
  models.
\newblock {\em EPL (Europhysics Letters)}, 82(5):57003, 2008.

\bibitem{Gull2011a}
Emanuel Gull, Andrew~J. Millis, Alexander~I. Lichtenstein, Alexey~N. Rubtsov,
  Matthias Troyer, and Philipp Werner.
\newblock Continuous-time monte carlo methods for quantum impurity models.
\newblock {\em Rev. Mod. Phys.}, 83(2):349, May 2011.

\bibitem{DGAintro}
K.~Held.
\newblock Dynamical vertex approximation.
\newblock {\em Dynamical Vertex Approximation [arXiv:1411.5191]}, 2014.
\newblock in Lecture Notes "Autumn School on Correlated Electrons. DMFT at 25:
  Infinite Dimensions", Reihe Modeling and Simulation, Vol. 4,
  Forschungszentrum Juelich GmbH (publisher), E. Pavarini, E. Koch, D.
  Vollhardt, and A. I. Lichtenstein (editors) [ISBN 978-3-89336-953-9].

\bibitem{DGA}
A.~Toschi, A.~A. Katanin, and K.~Held.
\newblock Dynamical vertex approximation; a step beyond dynamical mean-field
  theory.
\newblock {\em Phys Rev. B}, 75:045118, 2007.

\bibitem{DF}
A.~N. Rubtsov, M.~I. Katsnelson, A.~I. Lichtenstein, and A.~Georges.
\newblock Dual fermion approach to the two-dimensional hubbard model:
  Antiferromagnetic fluctuations and fermi arcs.
\newblock {\em Phys. Rev. B}, 79(4):045133, 2009.

\bibitem{1PI}
G.~Rohringer, A.~Toschi, H.~Hafermann, K.~Held, V.~I. Anisimov, and A.~A.
  Katanin.
\newblock One-particle irreducible functional approach: A route to diagrammatic
  extensions of the dynamical mean-field theory.
\newblock {\em Phys. Rev. B}, 88:115112, 2013.

\bibitem{RMPVertex}
G.~Rohringer, H.~Hafermann, A.~Toschi, A.~A. Katanin, A.~E. Antipov, M.~I.
  Katsnelson, A.~I. Lichtenstein, A.~N. Rubtsov, and K.~Held.
\newblock {Diagrammatic routes to non-local correlations beyond dynamical mean
  field theory}.
\newblock {\em ArXiv e-prints}, 2017.

\bibitem{Katanin2009}
A.~A. Katanin, A.~Toschi, and K.~Held.
\newblock Comparing pertinent effects of antiferromagnetic fluctuations in the
  two- and three-dimensional hubbard model.
\newblock {\em Phys. Rev. B}, 80:075104, Aug 2009.

\bibitem{Rubtsov2009}
A.~N. Rubtsov, M.~I. Katsnelson, A.~I. Lichtenstein, and A.~Georges.
\newblock Dual fermion approach to the two-dimensional hubbard model:
  Antiferromagnetic fluctuations and fermi arcs.
\newblock {\em Phys. Rev. B}, 79(4):045133, 2009.

\bibitem{Jung2010}
C.~Jung.
\newblock {\em Superperturbation theory for correlated fermions}.
\newblock PhD thesis, University of Hamburg, 2010.

\bibitem{Jung2011}
C.~Jung, A.~Wilhelm, H.~Hafermann, S.~Brener, and A.~Lichtenstein.
\newblock Superperturbation theory on the real axis.
\newblock {\em Annalen der Physik}, 523(8-9):706--714, 2011.

\bibitem{Taranto2014}
C.~Taranto, S.~Andergassen, J.~Bauer, K.~Held, A.~Katanin, W.~Metzner,
  G.~Rohringer, and A.~Toschi.
\newblock From infinite to two dimensions through the functional
  renormalization group.
\newblock {\em Phys. Rev. Lett.}, 112:196402, May 2014.

\bibitem{Schaefer2015-2}
T.~Sch\"afer, F.~Geles, D.~Rost, G.~Rohringer, E.~Arrigoni, K.~Held,
  N.~Bl\"umer, M.~Aichhorn, and A.~Toschi.
\newblock Fate of the false mott-hubbard transition in two dimensions.
\newblock {\em Phys. Rev. B}, 91:125109, Mar 2015.

\bibitem{Rohringer2011}
G.~Rohringer, A.~Toschi, A.~Katanin, and K.~Held.
\newblock Critical properties of the half-filled hubbard model in three
  dimensions.
\newblock {\em Phys. Rev. Lett.}, 107:256402, Dec 2011.

\bibitem{Antipov2014}
Andrey~E. Antipov, Emanuel Gull, and Stefan Kirchner.
\newblock Critical exponents of strongly correlated fermion systems from
  diagrammatic multiscale methods.
\newblock {\em Phys. Rev. Lett.}, 112:226401, Jun 2014.

\bibitem{Hirschmeier2015}
D.~Hirschmeier, H.~Hafermann, E.~Gull, A.~I. Lichtenstein, and A.~E. Antipov.
\newblock Mechanisms of finite-temperature magnetism in the three-dimensional
  hubbard model.
\newblock {\em Phys. Rev. B}, 92:144409, Oct 2015.

\bibitem{Schaefer2016}
T.~Sch\"afer, A.A. Katanin, K.~Held, and A.~Toschi.
\newblock Quantum criticality with a twist - interplay of correlations and kohn
  anomalies in three dimensions.
\newblock {\em preprint}, 2016.

\bibitem{Hafermann2009}
H.~Hafermann, G.~Li, A.~N. Rubtsov, M.~I. Katsnelson, A.I. Lichtenstein, and
  H.~Monien.
\newblock Efficient perturbation theory for quantum lattice models.
\newblock {\em Phys. Rev. Lett.}, 102:206401, 2009.

\bibitem{Ribic2017}
T.~Ribic, G.~Rohringer, and K.~Held.
\newblock Local correlation functions of arbitrary order for the
  falicov-kimball model.
\newblock {\em Phys. Rev. B}, 95:155130, 2017.

\bibitem{Gull2008}
E.~Gull, P.~Werner, X.~Wang, M.~Troyer, and A.~J. Millis.
\newblock Local order and the gapped phase of the hubbard model: A plaquette
  dynamical mean-field investigation.
\newblock {\em EPL (Europhysics Letters)}, 84(3):37009, 2008.

\bibitem{Hafermann2012}
Hartmut Hafermann, Kelly~R. Patton, and Philipp Werner.
\newblock Improved estimators for the self-energy and vertex function in
  hybridization-expansion continuous-time quantum monte carlo simulations.
\newblock {\em Phys. Rev. B}, 85:205106, May 2012.

\bibitem{Kaufmann2017}
J.~Kaufmann, P.~Gunacker, and K.~Held.
\newblock High-frequency asymptotics of multi-orbital vertex functions in
  continuous-time quantum monte carlo.
\newblock {\em preprint}, 2017.

\bibitem{JPSJ-XXX}
A.~{Galler}, J.~{Kaufmann}, P.~{Gunacker}, P.~{Thunstr{\"o}m}, J.~M. {Tomczak},
  and K.~{Held}.
\newblock {Towards ab initio calculations with the dynamical vertex
  approximation}.
\newblock {\em ArXiv e-prints}, September 2017.

\bibitem{Brener2008}
S.~Brener, H.~Hafermann, A.~N. Rubtsov, M.~I. Katsnelson, and A.~I.
  Lichtenstein.
\newblock Dual fermion approach to susceptibility of correlated lattice
  fermions.
\newblock {\em Phys. Rev. B}, 77:195105, May 2008.

\bibitem{sixpoint}
A.~A. Katanin.
\newblock The effect of six-point one-particle reducible local interactions in
  the dual fermion approach.
\newblock {\em Journal of Physics A: Mathematical and Theoretical},
  46(4):045002, 2013.

\bibitem{Note1}
The presented DF results are without self-consistency. However, for the
  parameters considered, imposing an inner self-energy self-consistency (not
  shown) leads only to minor modifications for $U=1$ and reduces both two- and
  three-particle corrections to about half their values for $U=2$. A closer
  investigation should also include an outer self-consistency with an update of
  the vertex and local problem, but is beyond the scope of the present paper.

\bibitem{RohringerToschi2016}
G.~Rohringer and A.~Toschi.
\newblock Impact of nonlocal correlations over different energy scales: A
  dynamical vertex approximation study.
\newblock {\em Phys. Rev. B}, 94:125144, Sep 2016.

\bibitem{HubbDGA}
T~Sch{\"a}fer, A~Toschi, and K~Held.
\newblock Dynamical vertex approximation for the two-dimensional hubbard model.
\newblock {\em Journal of Magnetism and Magnetic Materials}, 1506(05706), 2015.

\bibitem{Matplotlib}
John~D. Hunter.
\newblock Matplotlib: A 2d graphics environment.
\newblock {\em Computing In Science \& Engineering}, 9, May-Jun 2007.

\bibitem{w2d1}
Nicolaus Parragh, Alessandro Toschi, Karsten Held, and Giorgio Sangiovanni.
\newblock Conserved quantities of ${SU(2)}$-invariant interactions for
  correlated fermions and the advantages for quantum monte carlo simulations.
\newblock {\em Phys. Rev. B}, 58:155158, 2012.

\bibitem{w2d2}
Markus~Wallerberger et~al.
\newblock in preparation.

\bibitem{ALPSnew}
A.~Gaenko, {A. E.} Antipov, G.~Carcassi, T.~Chen, X.~Chen, Q.~Dong, L.~Gamper,
  J.~Gukelberger, R.~Igarashi, S.~Iskakov, M.~K{\"o}nz, {J. P.F.} LeBlanc,
  R.~Levy, {P. N.} Ma, {J. E.} Paki, H.~Shinaoka, S.~Todo, M.~Troyer, and
  E.~Gull.
\newblock Updated core libraries of the alps project.
\newblock {\em Comput. Phys. Commun.}, 213:235--251, 4 2017.

\bibitem{OpenDF}
Andrey~E. Antipov, James~P.F. LeBlanc, and Emanuel Gull.
\newblock Opendf - an implementation of the dual fermion method for strongly
  correlated systems.
\newblock {\em Physics Procedia}, 68:43 -- 51, 2015.
\newblock Proceedings of the 28th Workshop on Computer Simulation Studies in
  Condensed Matter Physics (CSP2015).

\bibitem{Note2}
In the denominator, the $1$ in the expansion $\protect \qopname \relax
  o{exp}^{\protect \mathcal {S}_{dual}^{k}}=\left (1+\protect \mathcal
  {S}_{dual}^{k}\right )$ vanishes when integrating. The remaining term
  $\protect \mathcal {S}_{dual}^{k}$ can be rewritten according to Eq.~(\ref
  {GrassMann}).

\end{thebibliography}

\end{document}